\documentclass[12pt,a4paper]{article}

\usepackage{ccaption}
\usepackage{dcolumn}
\usepackage{amsmath}
\usepackage{amsfonts}
\usepackage{amssymb}
\usepackage{amsthm}
\usepackage{graphicx}
\usepackage{setspace}

\newcommand{\E}{\mathbb{E}}

\newcommand \Tstyle {\footnotesize \captionnamefont{\footnotesize}
\captiontitlefont{\footnotesize}
\captionstyle{\flushleftright}}

\title{Hedging without sweat: a genetic programming approach}

\author{Terje Lensberg$^{a}$ and Klaus Reiner Schenk-Hopp{\'e}$^{a,b}$}

\date{May 14, 2013\vspace*{-2mm}}

\begin{document}


\maketitle

\renewcommand{\thefootnote}{\alph{footnote}}

\footnotetext[1]{Department of Finance, NHH--School of Economics, Bergen, Norway.
\newline \hspace*{3.1ex}$\rm ^{b}$Leeds University Business School and School of Mathematics, University of Leeds, United Kingdom.
\newline  \hspace*{3.1ex}
Email addresses: terje.lensberg@nhh.no; k.r.schenk-hoppe@leeds.ac.uk.
\newline  \hspace*{3.1ex} This paper was completed during a visit to the Hausdorff Research Institute for Mathematics at the University of Bonn in the framework of the Trimester Program \textit{Stochastic Dynamics in Economics and Finance}. Financial support by the European Commission under the Marie Curie Fellowship Programme (grant agreement PIEF-GA-2010-274454) is gratefully acknowledged.}

\renewcommand{\thefootnote}{\arabic{footnote}}
\setcounter{footnote}{0}

\doublespacing

\begin{abstract}
\noindent Hedging in the presence of transaction costs leads to complex optimization problems. These problems typically lack closed-form solutions, and their implementation relies on numerical methods that provide hedging strategies for specific parameter values. In this paper we use a genetic programming algorithm to derive explicit formulas for near-optimal hedging strategies under nonlinear transaction costs. The strategies are valid over a large range of parameter values and require no information about the structure of the optimal hedging strategy.
\end{abstract}

\noindent \textit{Keywords:} Hedging, transaction costs, closed-form approximations, genetic programming.

\noindent \textit{JEL classification:} G13.

\section{Introduction}\label{sec:intro}

Transaction costs preclude perfect replication of contingent claims and introduce a tradeoff between the risk and return to hedging. A standard approach to dealing with this tradeoff is to use utility functions for evaluating hedging errors (Hodges and Neuberger 1989) and the utility indifference principle for pricing options (Davis et al.\ 1993). In contrast to the elegance of Black-Scholes delta hedging, these models typically lack explicit solutions and require powerful numerical methods to obtain case-by-case approximations of optimal hedges.

In some situations the particular structure of the optimal hedging strategy is known, which facilitates numerical analysis. One example is the Black-Scholes model with proportional transaction costs, where the hedging strategy of a trader with CARA utility function is defined by a no-trade region (Davis et al.\ 1993). When the hedger's stock holdings are outside this region, she carries out a trade that brings the stock position onto the nearest boundary point. Numerical methods that approximate value functions have proved successful in this model, see, e.g., Monoyios (2004). A different approach is presented in Zakamouline (2006) who derives analytic approximations of the two boundaries of the no-trade region. Motivated by asymptotic results in Whalley and Wilmott (1997), his approach combines a clever guess of a parametric functional form with the estimation of coefficients as functions of model parameters. These estimations, however, require a numerical approximation of the optimal hedging strategy.

In this paper we propose an alternative approach to derive explicit hedging strategies for general nonlinear transaction costs. The approach is based on genetic programming which offers several advantages over classical numerical approaches. First, by construction it is constrained to output closed-form expressions that are functions of the model parameters. Second, it delivers efficient results at low cost by solving the two tasks of finding an optimal hedging strategy and a closed-form representation for it in one integrated process. Third, no assumption is made on the functional form of the approximation. Genetic programming therefore avoids the disadvantage of ex post calibration methods and, at the same time, offers a technique that deals with cases in which there is no a priori information on the structure of the optimal solution. The closed-form approximations of hedging strategies derived with this heuristic method can be directly tested in different market situations to verify their efficiency before their integration into an automated trading system. The hard work of obtaining these hedging strategies is left to the computer. Once this is done, hedging is without sweat.

In the example considered in this paper we determine closed-form expressions for the hedging strategy of an option writer with CARA utility function under proportional and quadratic transaction costs. The price of the underlying follows a geometric Brownian motion. Solutions to this problem with a large range of parameter values are derived at various requirements on the simplicity of hedging strategies, ranging from full-fledged computer programs to short analytic functions. Parameters include volatility, interest rate, strike price of the European call, and transaction cost parameters. Information available to the hedger consists of stock price, Black-Scholes delta and gamma, time to maturity, and current portfolio holdings, in addition to the parameter values which are random but constant until maturity of an option.

Our method produces better results than Zakamouline (2006) when transaction costs are proportional. Under quadratic transaction costs we find simple near-optimal linear trading strategies that consist of a trade intensity and a no-trade reference solution. The no-trade path can be interpreted as a modified Black-Scholes delta and the trade intensity as a modified gamma. In summary, the paper demonstrates the practical use of genetic programming in deriving hedges under nonlinear transaction costs.

\section{Model}\label{sec:model}

We consider the hedging problem faced by a risk-averse writer of a European call option. The option writer can trade in the underlying whose price follows a geometric Brownian motion
\begin{equation}\label{eq:BM}
    dS(t) = \mu S(t) dt + \sigma S(t) dW(t)
\end{equation}
with constant drift $\mu$ and volatility $\sigma$. She also has access to a money account that pays interest at constant rate $r$. We set $\mu = r$ to obtain the risk-neutral measure. The option writer can trade only at dates $t_n = n \, \delta t$ with $n=0,...,N$, and $T = t_N$ the expiry date of the option.

Trading incurs transaction costs. We assume that the total transaction cost of buying or selling $x$ shares at the current price $S$ is given by
\begin{equation}\label{eq:totalcost}
    (\lambda + \beta S |x|) S |x|
\end{equation}
with $\lambda \geq 0$ and $\beta \geq 0$. It is deducted from the trader's money account at the time a transaction takes place. The transaction cost has two components: (a) A proportional cost $\lambda$ which arises in markets with a bid-ask spread of size $2 \lambda S$ and mid point given by the stock price $S$; and (b) a nonlinear cost which arises in markets when a trade can `walk up' an order book, i.e., obtains worse prices the larger the trade.

\begin{table}[ht]
  \centering
\Tstyle
  \caption{Parameter values.}
\label{table:parameters}
\begin{tabular}{lcl}
\\[-1.8ex]\hline
\hline \\[-1.8ex]
  Risk aversion &  $\gamma$ &  0.5\\
  Proportional cost &  $\lambda$ &  10bp - 200bp  \\
  Quadratic cost &  $\beta$ &  5bp - 100bp  \\
  Volatility &  $\sigma$ &  uniformly drawn from $[10\%, 40\%]$ \\
  Interest rate &  $r$ &  uniformly drawn from $[1\%, 10\%]$ \\
  Strike price &  $K$ &  uniformly drawn from $[(1-\sigma)100 , (1+\sigma)100]$ \\
   & &  where the spot price $S(t_0) = 100$\\
  Maturity &  $T$ &  3 months \\
  Hedge frequency &   $\delta t$ &  $1/264$ years (one day) \\
\\[-1.8ex]\hline
\hline \\[-1.8ex]
\end{tabular}
\end{table}

Trading strategies are real-valued functions of the form
\begin{equation}\label{eq:strat}
\phi(t_n) = \phi_{\theta}(t_n, S(t_n), \Delta(t_n), \Gamma(t_n), x(t_{n-1}))
\end{equation}
where $\phi(t_n)$ is the number of shares bought or sold at time $t_n$, and $\theta = (\sigma,r,K,\lambda,\beta)$ is a parameter vector which is constant over the lifetime of the option, cf. Table~\ref{table:parameters}. The function $\phi$ can depend on $\theta$, as well as time $t_n$, stock price $S(t_n)$, the Black-Scholes greeks $\Delta(t_n)$ and $\Gamma(t_n)$, and the number of shares $x(t_{n-1})$ held after trading in period $t_{n-1}$. In our application, this function will be represented by a computer program.

Share holdings and money market account evolve as
\begin{equation*}
     \begin{aligned}
x_{\phi}(t_{n}) & = x_{\phi}(t_{n-1}) + \phi(t_n)\\
y_{\phi}(t_{n}) & = \exp(r \delta t) y_{\phi}(t_{n-1}) - S(t_n) \phi(t_n) - (\lambda + \beta S(t_n) |\phi(t_n)|) \, S(t_n) |\phi(t_n)|
     \end{aligned}
\end{equation*}
with initial portfolio $(x_{\phi}(t_{-1}), y_{\phi}(t_{-1}))$. We make the usual assumption that there are no transaction costs at maturity. The terminal wealth of an option writer who sold one European call option with strike price $K$ at time $t_0 = 0$ is therefore given by
\begin{equation}\label{eq:termw}
w_{\phi}(T) = S(T) x_{\phi}(T) + y_{\phi}(T) - [S(T) - K]^+.
\end{equation}

The option writer's objective is to maximize expected utility $\E [u(w_{\phi}(T))]$ from terminal wealth, where the expectation is taken over the probability measure defined by \eqref{eq:BM} on the space of sample paths.  Given two hedging strategies $\phi'$ and $\phi''$, we measure their relative performance for an option $\theta = (\sigma, r, K, \lambda, \beta)$ by the difference in certainty equivalents
\begin{equation}\label{eq:dp}
\Pi(\phi_{\theta}', \phi_{\theta}'') := u^{-1} \left(\E  [u(w_{\phi_{\theta}'}(T))]\right) - u^{-1}\left(\E [u(w_{\phi_{\theta}''}(T))]\right)
\end{equation}
where $\E [u(w_{\phi_{\theta}}(T))]$ is the expected utility of terminal wealth when hedging strategy $\phi$ is applied to option $\theta$. $\Pi(\phi_{\theta}', \phi_{\theta}'')$ is the monetary gain from using strategy $\phi'$ instead of $\phi''$ to hedge option $\theta$, and $\phi'$ performs better than $\phi''$, the larger is $\Pi(\phi_{\theta}', \phi_{\theta}'')$.

We will assume that the option writer has CARA utility
$$
u(w) = -\exp(-\gamma \, w), \qquad \gamma > 0.
$$
For this class of utility functions, $\Pi$ is independent of initial money holdings which therefore can be set to zero for convenience. It also allows to determine option prices through the indifference principle: the option premium is determined by the amount of money that when paid to the option writer gives the same utility as not writing the option and forgoing the premium. If the initial endowment is $(0,0)$ (which gives utility -1 if no option is written), then the price is
$$
p_{\theta} = - \exp(- r T) \, u^{-1}\left( \E [u(w_{\phi_{\theta}}(T))] \right) = (1/\gamma) \exp(- r T) \, \ln \left( \E \exp(-\gamma w_{\phi_{\theta}}(T)) \right)
$$
for a hedger with strategy $\phi_{\theta}$. This indifference price will typically depend on the initial share endowment through its impact on (a) the trading strategy of the option writer and (b) the opportunity cost which is the utility of an investor with no position in the option.

\section{Genetic programming}\label{sec:gp}

Solving the hedging problem described in Section~\ref{sec:model} is a straightforward application of genetic programming (Koza 1992). We use a parallel steady-state algorithm with tournament selection, as illustrated in Table~\ref{table:gpalgo}.

\begin{table}[ht]
 \centering
\Tstyle
  \caption{Parallel GP algorithm with many worker processes and a master process. The workers search for increasingly better hedging strategies. The master coordinates and collects results. Performance is measured as average realized utility across random sets of options.}
  \label{table:gpalgo}
\begin{tabular}{ll}
\\[-1.8ex]\hline
\hline \\[-1.8ex]
Worker processes & Master process \\
\hline \\[-1.8ex]
\begin{minipage}[t]{0.47\linewidth}
Randomly generate 250 programs\\
For iteration 1,...,250:\\
\hspace*{1.5ex}  Randomly generate 20,000 options\\
\hspace*{1.5ex}  Do at least 125 tournaments:\\
\hspace*{3ex}    Select and evaluate 4 random programs\\
\hspace*{3ex}     Rank them by performance\\
\hspace*{3ex}     Replace worst 2 by copies of best 2\\
\hspace*{3ex}     Cross and mutate the copies\\
\hspace*{1.5ex}   End Do\\
\hspace*{1.5ex}   Send best program to master\\
\hspace*{1.5ex}   Exchange good programs with neighbors\\
End For
\end{minipage}
&
\begin{minipage}[t]{0.47\linewidth}
Randomly generate 100,000 options\\
For iteration 1,...,250:\\
\hspace*{1.5ex}  Wait for all workers to do at least 125\\
\hspace*{4.5ex}   tournaments\\
\hspace*{1.5ex}    Tell workers to send programs and start\\
\hspace*{4.5ex}     the next iteration\\
\hspace*{1.5ex}  Receive programs from workers\\
\hspace*{1.5ex}  Evaluate programs\\
\hspace*{1.5ex}  Save best program and report results\\
End For
\end{minipage} \\ \\[-2mm]
\hline
\hline
\end{tabular}
\end{table}

The algorithm uses some 100-500 autonomous sub-populations of hedging strategies deployed on separate worker processors arranged in a circle topology. From time to time, the worker processes send their best program to a master process; exchange good candidate solutions with their neighbors; and generate a new set of test options. The master process uses a separate large fixed set of options to identify the current globally best strategy. A byte code program representation is used for genetic recombination (crossover and mutation), and a machine code representation (Nordin 1997) is used for fast computation of hedging decisions. A simple built-in compiler translates byte code to machine code. Our software also contains a byte code disassembler whose output can be processed by a C compiler, or analyzed with Maxima or similar tools for symbolic math manipulation.

\section{Proportional transaction costs}\label{sec:prop}

We first apply the genetic programming (GP) method to the model with proportional transaction costs (\eqref{eq:totalcost} with $\lambda > 0$ and $\beta = 0$). Two different treatments will be applied. Trading strategies are first evolved using the GP approach without any information about the structure of the optimal hedge strategies. We choose to ignore the knowledge that under proportional transaction costs the optimal strategy is defined by a no-trade region because under other cost structures such a priori information is typically not available. In a second treatment, we derive a simple analytic closed-form expression to approximate the no-trade region.

\subsection{GP-model without structural information}\label{sec:prop_constr}

The structure of the hedging strategies is given by \eqref{eq:strat}. Strategies are represented as computer programs whose operators include arithmetic expressions as well as \textit{min}, \textit{max}, conditional assignments, and forward jumps (conditional and unconditional). The output from a program is interpreted as a decision of how much stock to buy or sell given the current combination of option parameters and variables. The programs can be used as is in an automated trading application, although they may be difficult to analyze due to the occurrence of conditional jumps and assignments.

\begin{figure}
\begin{center}
\begin{tabular}{lr}
\includegraphics[width = .50\textwidth]{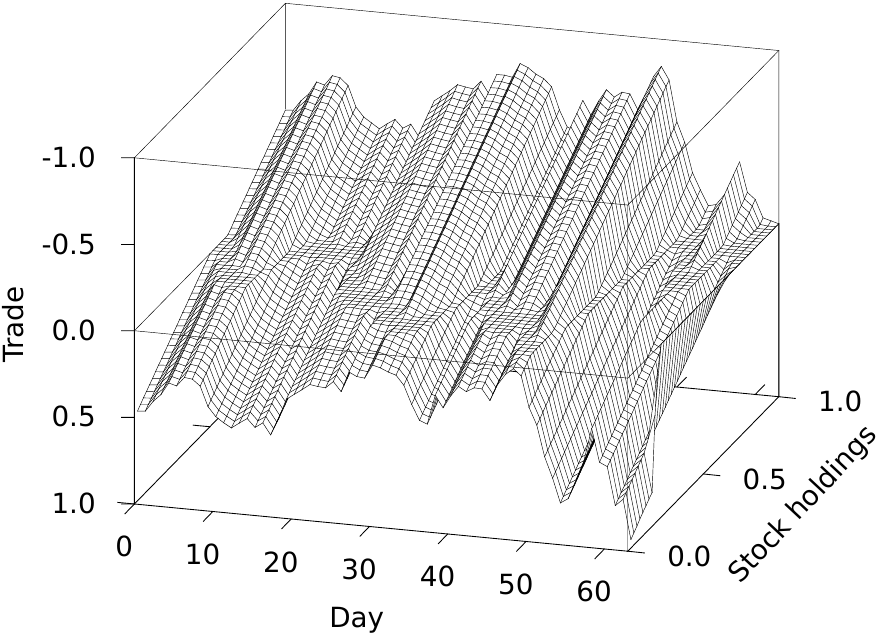} & \includegraphics[width = .45\textwidth]{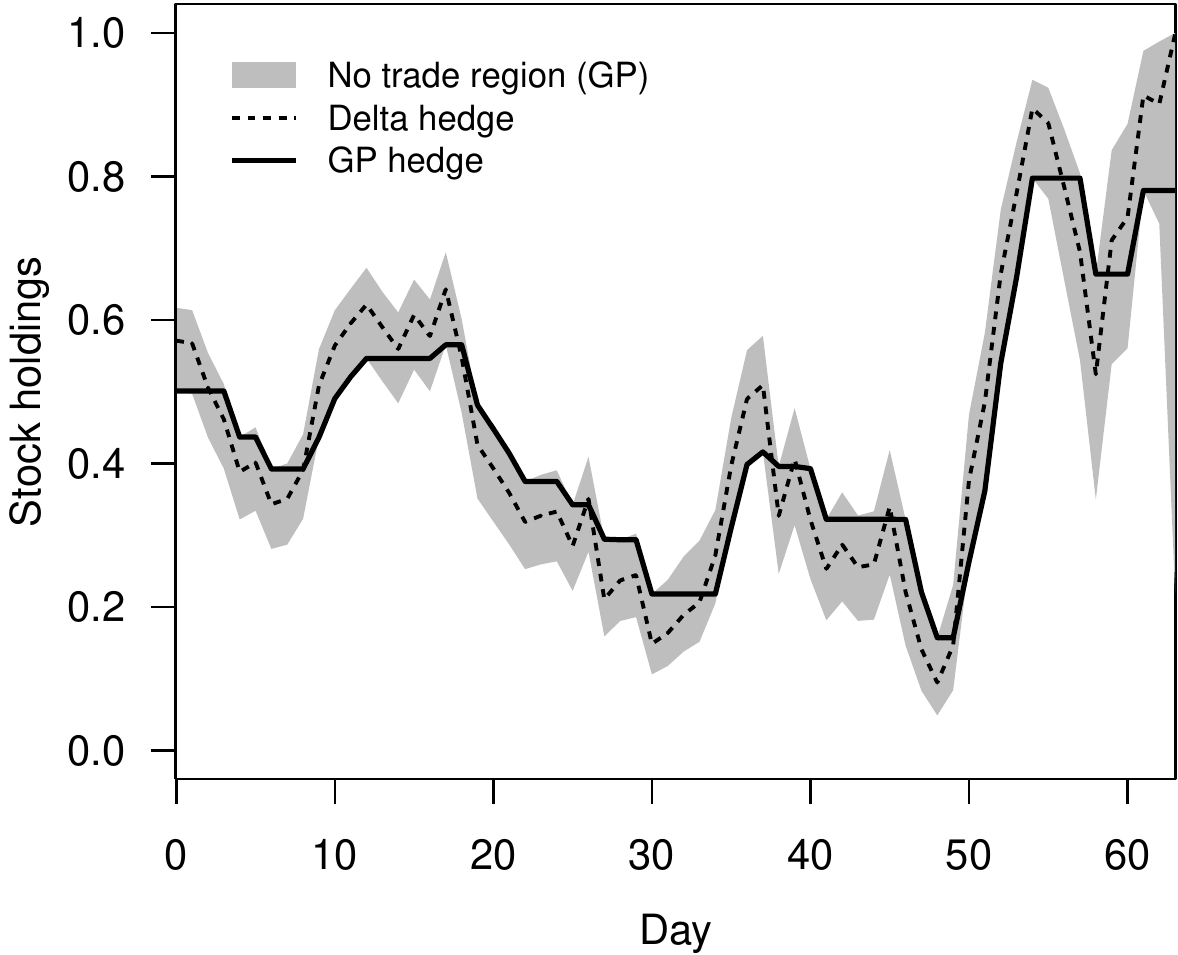} \\
\end{tabular}
\caption{Proportional costs. Left panel: Trade volume by day and stock position. Right panel: Time series of hedging strategies' stock position. GP hedge (bold black line), no-trade region of the GP hedge (shaded area), and Black-Scholes delta (dotted line). Parameter values are $\theta = (\sigma,r,K,\lambda,\beta) = (17.38\%,3.17\%,99.6,20\mbox{bp},0)$ and the initial endowment is $(0,0)$.} \label{fig:tradingPC}
\end{center}
\end{figure}

Our approximation of the optimal hedging strategy is illustrated in Figure~\ref{fig:tradingPC} for an example option and stock price path. The left panel depicts trade volumes for each day and each stock position in $[0,1]$. It reveals the existence of a no-trade region such that the trade volume for stock positions outside this region brings the hedger's position onto its closest boundary. We conclude that the GP algorithm is able to identify the structure of the optimal hedging strategy.

The right panel depicts a time-series of the GP hedger's stock position, the corresponding no-trade region and, as reference, the Black-Scholes delta hedge. The no-trade region is extracted from the GP hedging strategy by determining, at each point in time, all stock positions at which the trade volume is zero. The figure shows that the GP hedge position does indeed move as if constrained by the boundaries of the no-trade region.

We now turn to the issue whether the approximation produced with the GP approach is a good one. As a benchmark  we use the trading strategy derived from Zakamouline's (2006) approximation of the no-trade region. The Zakamouline hedge is also a closed-form approximation, which provides a level playing field, and, according to the tests in that paper, the best available.

\begin{table}[ht]
\vspace*{-13mm}
\centering
\Tstyle
  \caption{Comparison of the GP and Zakamouline hedging strategies for proportional transaction costs and a CARA utility function with risk aversion $\gamma = 0.5$. The statistical analysis is carried out on a data set which is generated as follows. We create an orthogonal and equidistant grid of $10^4$ points on the set of option parameters: Strike $K \in [91,109]$, volatility $\sigma \in [11.5\%,38.5\%]$, interest rate $r \in [1.5\%,9.5\%]$ and transaction costs $\lambda \in [0.15\%,1.95\%]$. Moneyness is defined as $\ln(S(t_0)/K)$. Its absolute value is used as a measure of the distance between the initial stock price and the strike. For each point $\theta = (\sigma, r, K, \lambda, 0)$ in the grid, we simulate 100,000 random price paths and compute the relative performance measure $\Pi(GP_\theta,Z_\theta)$ defined in (\ref{eq:dp}) for the GP and Zakamouline hedging strategies. Using these 10,000 observations, we then regress the performance measure on standardized option parameters. Differences in performance are measured in cents per option. Standard errors in parentheses.}
  \label{table:utility}
  \begin{tabular}{@{\extracolsep{5pt}}lD{.}{.}{-3} D{.}{.}{-3} D{.}{.}{-3} D{.}{.}{-3} }
\\[-1.8ex]\hline
\hline \\[-1.8ex]
 & \multicolumn{4}{c}{\textit{Dependent variable:}} \\
\cline{2-5}
\\[-1.8ex] & \multicolumn{4}{c}{$100 \times \Pi(GP,Z)$} \\
\\[-1.8ex] & \multicolumn{1}{c}{(1)} & \multicolumn{1}{c}{(2)} & \multicolumn{1}{c}{(3)} & \multicolumn{1}{c}{(4)}\\
\hline \\[-1.8ex]
 Constant & 4.601 & 4.601 & 4.601 & 4.601 \\
  & (0.034) & (0.030) & (0.029) & (0.029) \\[1mm]
 Transaction cost & 4.478 & 9.033 & 9.033 & 9.033 \\
  & (0.034) & (0.091) & (0.090) & (0.090) \\[1mm]
 Volatility & -1.968 & 0.901 & 0.901 & 0.901 \\
  & (0.034) & (0.062) & (0.061) & (0.061) \\[1mm]
 Transaction cost $\times$ Volatility &  & -5.607 & -5.607 & -5.607 \\
  &  & (0.106) & (0.105) & (0.105) \\[1mm]
 abs(Moneyness) &  &  & -0.403 & -0.403 \\
  &  &  & (0.029) & (0.029) \\ [1mm]
 Interest rate &  &  &  & -0.080 \\
  &  &  &  & (0.029) \\[1mm]
\hline \\[-1.8ex]
Observations & \multicolumn{1}{c}{\hspace{2mm} $10,000$} \\
Adjusted R$^{2}$ & 0.680 & 0.750 & 0.755 & 0.755 \\
\hline
\hline \\[-1.8ex]
\multicolumn{4}{l}{\hspace{-3.25mm} \textit{Note:} All coefficients are significantly different from zero at the 1\% level.}  \\
\normalsize
\end{tabular}
\end{table}

Table~\ref{table:utility} provides a detailed comparison of the performance of the GP and the Zakamouline hedging strategies. The statistical analysis is carried out by applying the performance measure $\Pi$ defined in  defined in (\ref{eq:dp}) to the GP and Zakamouline hedges on 10,000 parameter vectors $\theta$. These values are contained in the sets on which both hedging strategies are defined to ensure a fair comparison.

On average, the GP hedge outperforms the Zakamouline hedge by 4.6 cents per option. This amounts to approximately 0.5\% of the mean Black-Scholes option price, which is 850 cents. (All explanatory variables are standardized, so the constant term in the regression is the unconditional mean of the dependent variable.) The best performance of GP over the Zakamouline hedge is 45.9 cents while the worst performance is -3.3 cents. The GP hedge performs better in 79\% of the 10,000 cases. The difference in performance varies across cases in a systematic way. Model (1) shows that the GP hedging strategy is the better, the higher the transaction cost $\lambda$ and the lower the volatility $\sigma$. In model (2) we add the cross-term of transaction costs and volatility. The table shows that the relative performance of the GP hedge is best for high transaction costs and low volatility, and worst for the opposite parameter configuration. When transaction costs are low and volatility high, the Zakamouline and the optimal hedges are both quite close to the Black-Scholes delta, which leaves little scope for improvement.

The effect of option moneyness on the difference in performance is tested in model (3). Moneyness adds very little in terms of explanatory power, but the GP hedge is slightly better for options at the money, and slightly worse for options far into or far out of the money. This is as one might expect since the latter cases are associated with low trade volumes which reduces the potential for improving performance.  Finally, the interest rate has a significant (but economically negligible) effect on the difference in performance as documented in model (4).

In most cases our hedging strategy outperforms Zakamouline (2006) which, by itself, is better than other closed-form approximations. Reasons for this result are that (i) Zakamouline's model produces more tail risk because it is calibrated on data generated by hedges that are derived using a quadratic approximation of the negative exponential utility function; (ii) there are better functional forms for analytic approximations than log-linear (demonstrated in the next section); and (iii) there seems to be a benefit in not including risk aversion as a parameter. Observation (iii) seems to contradict Zakamouline's (2006, p.\ 441) argument but it is correct in light of (ii).

\subsection{GP-model with no-trade region}\label{sec:prop_notrade}

This section applies the GP methodology to develop a simple analytic approximation to the no-trade region under proportional costs. To this end, we evolve computer programs with two outputs: the lower and the upper bound of the no-trade region. Jumps and conditional assignments are excluded from the list of feasible instructions, thereby forcing the computer programs to represent `simple' analytic functions. As these functions can consist of up to 256 operations, they can nevertheless be quite complex. To favor simplicity over complexity, one can modify the fitness criterion used by the GP-algorithm by introducing a small penalty for program length. Although the outcome of this method is not unique, one consistently obtains hedging strategies that perform only slightly worse than the best attainable.

We find that the best strategy in this class outperforms the best unconstrained strategy of Section~\ref{sec:prop_constr} by 0.6 cents on average. Its best improvement over the Zakamouline hedge is 46.2 cents; the worst is -0.9 cents, and it performs better in 90\% of the 10,000 cases. We also find that more complexity adds surprisingly little to the average performance of hedges. Consider, for instance, the analytic approximation of the no-trade region $[\Delta_t-L_t,\Delta_t+U_t]$ given by
\begin{equation*}
     \begin{aligned}
L_t & = \Gamma_t + {\Delta_t }^{2}  \left( \Gamma_t + A_t  \right)\\
U_t & = \Gamma_t + \left( 1-\Delta_t \right)  A_t
     \end{aligned}
\end{equation*}
with
$$
A_t = 6 \, \frac{\lambda \tau \left( K +\frac{1.117573}{\sigma \tau}\right) + \sigma^2 }{\left( 2 \tau + \lambda \right)  \sigma S_t +\frac{1.117573}{\sigma \tau S_t \Gamma_t}}+\frac{\lambda \left( 1 - 4  \tau\right)  \left( K + \frac{1.117573}{\sigma \tau}\right) }{\sigma S_t}
$$
and $\tau = T-t$. This strategy produces hedges that, on average, perform as well as the best unconstrained strategy of Section~\ref{sec:prop_constr}, and only 0.5 cents worse than the best strategy in its class.

This analytic approximation of the no-trade region is obtained by (a) evolving 250 hedging strategies without imposing a penalty and (b) restarting the GP algorithm with these `candidate strategies' and adding a small penalty proportional to the combined length of the analytic expressions for the upper and lower bounds. The approximation is quite robust in the sense that its structure barely changes over more than 100 of the 250 iterations with the GP algorithm, although the constants vary a little.

It might seem surprising that the interest rate $r$ does not enter the expressions for the no-trade bounds other than via the option greeks. However, as also found in Zakamouline (2006, Tables 2 and 3), the direct effect of the interest rate on the no-trade bounds is very weak. Our result shows that interest rates in the range considered here are negligible for practical purposes.

\section{Quadratic transaction costs}\label{sec:quad}

We next consider hedging under quadratic transaction costs where the cost per dollar of trade increases with the total value of a trade. The choice of quadratic over other nonlinear cost structures is for simplicity of presentation but also leads to some interesting insights.

The specification can be justified as an approximation to the liquidity cost function used in \c{C}etin and Rogers (2007) as well as the short-term trade impact in limit order markets specified in Malo and Pennanen (2012). Using a second-order Taylor approximation of the exponential function, one finds that the total amount needed to buy, resp.\ the total amount received when selling, is
$$
\frac{\exp(\alpha S x) - 1}{\alpha} \approx \frac{1 + \alpha S x + (\alpha^2/2) S^2 x^2 - 1}{\alpha} =  S x + (\alpha/2) (S x)^2.
$$
The total transaction cost incurred is therefore $(\alpha/2) (S x)^2$ which, after replacing $\alpha$ by $2\beta$, corresponds to \eqref{eq:totalcost} with $\lambda = 0$. The same result is obtained for the cost function $S[\exp(\alpha x) - 1]/\alpha$ used in \c{C}etin and Rogers's (2007, Sect.\ 6).

\subsection{General GP-model}\label{sec:GPqcg}

We apply the GP method in its general form, allowing for jumps and conditional assignments, because there is no information on the structure of the optimal hedging strategy. The trading strategy will be of the form \eqref{eq:strat}. Buying or selling $x$ shares at market price $S$ costs $\beta (S x)^2$ which is deducted from the money account at the time of trade.

\begin{figure}
\begin{center}
\begin{tabular}{lr}
\includegraphics[width = .50\textwidth]{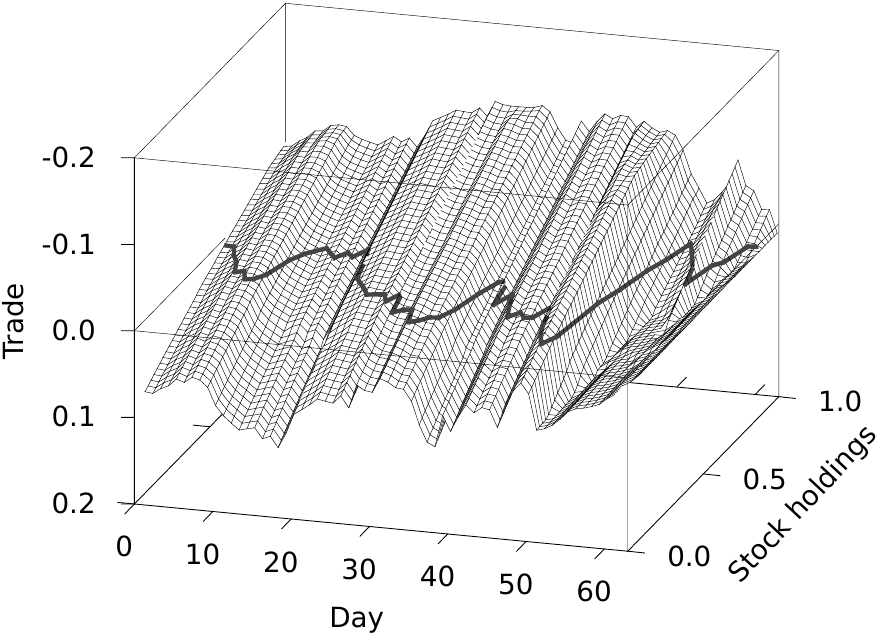} & \includegraphics[width = .45\textwidth]{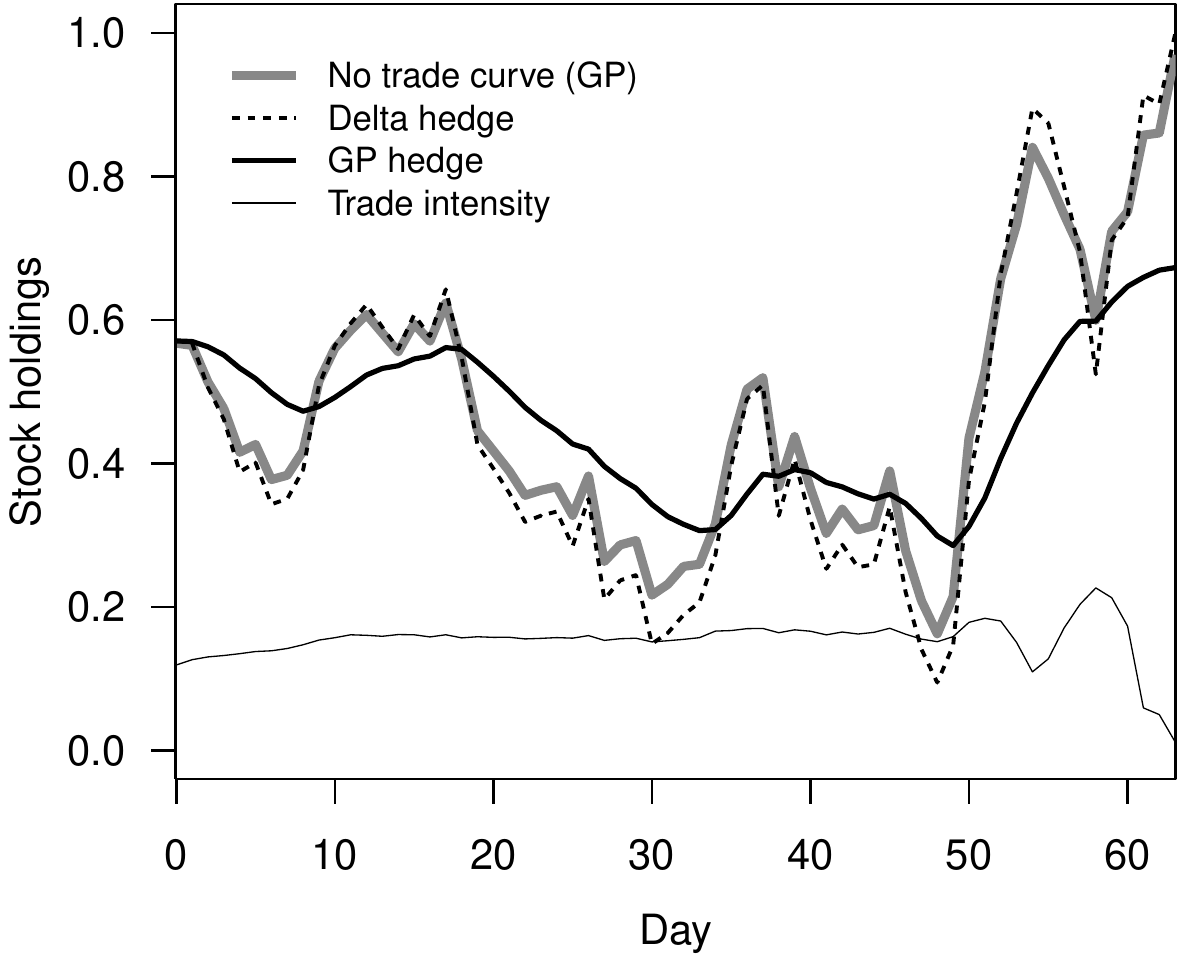} \\
\end{tabular}
\caption{Quadratic costs. Left panel: Trade volume by day and stock position. Right panel: Time series of GP hedge's stock position (bold black line), no-trade region of the GP hedge (grey line), trading intensity (black line) and, for reference, Black-Scholes delta (dotted line). Parameter values are $\theta = (\sigma,r,K,\lambda,\beta) = (17.38\%,3.17\%,99.6,0,10\mbox{bp})$ and the initial endowment is $(\Delta_0,0)$.}\label{fig:tradingQC}
\end{center}
\end{figure}

Figure~\ref{fig:tradingQC} provides an illustration of our results for an example option and stock price path. As in Figure~\ref{fig:tradingPC}, the left panel depicts trade volumes for each day and each stock position in $[0,1]$
for a typical trading strategy evolved with the GP algorithm, but in the this case with quadratic transaction costs. The left panel shows that the trade volume is approximately a linear function of the stock position for most stock positions and most periods in time. Intersecting the graph with the horizontal plane at zero on the vertical axis, one obtains a curve through combinations of time and stock positions where there is no trade, and such that for other nearby points on the graph, the trade volume on the corresponding day is a constant fraction of the difference between the corresponding no-trade position and stock position. These observations suggest that linear trading strategies will provide good performance under quadratic transaction costs.

The right-hand panel of Figure~\ref{fig:tradingQC} shows the projection of the no-trade curve (grey) along with the GP hedge (bold black), and the delta hedge (dotted). Also shown is the trade intensity (black), defined as trade volume divided by the absolute difference between the current stock position and the no-trade stock position. The latter closely tracks the Black-Scholes delta. The hedging strategy adjusts the stock position towards the no-trade stock position, but this adjustment is quite slow, as the trade intensity is typically less than 20\%. We therefore observe large deviations between actual stock holdings and the no-trade curve. Closer to maturity the trade intensity is more volatile but then declines quickly (resembling the widening of the no-trade region under proportional costs). This general pattern of sluggish adjustment leads to substantial savings in transaction costs compared to the Black-Scholes delta.

\subsection{Linear GP-model}

To explore the performance of linear hedging strategies under quadratic costs, we apply the GP algorithm to trading strategies of the form
\begin{equation}\label{eq:stratQC}
\phi(t_n) = \epsilon(t_n) \cdot \left[\bar{x}(t_n) - x(t_{n-1})\right]
\end{equation}
where $\bar{x}(t_n)$ is the no-trade stock position and $\epsilon(t_n)$ is the trading intensity, and both are analytic expressions. The functions are constrained to be independent of stock holdings but are allowed to depend on time, greeks, and the parameter vector $\theta$ which are suppressed in \eqref{eq:stratQC}.

As in the proportional cost case, making use of information about the (approximate) structure of the optimal hedge leads to strategies that are slightly better than those obtained without such information. In the current case, the mean improvement in performance amounts to 0.3 cents per option. This improvement is of the same order of magnitude as in the proportional cost case, where the a priori information was known to be correct. We therefore conjecture that linear strategies of the form (\ref{eq:stratQC}) are near-optimal for hedgers with CARA utility functions and quadratic transaction costs when the initial stock position is not too far from its desired level.

The following approximation was obtained by means of the same simplification approach as in Section~\ref{sec:prop_notrade}. It produces hedging decisions that perform as well as the best strategy in its class. For a given date $t$, the no-trade stock position $\bar{x}(t)$ and the trading intensity $\epsilon(t)$ are given by
\begin{equation*}
     \begin{aligned}
\bar{x}(t) & = \Delta_t - 0.6 (1-\tau)  \Gamma_t \left( \sigma + 25.19 \beta \right)   \left( \Delta_t + S_t - K - 2 \sigma - 0.8885\right) + 0.0034\\
\epsilon(t)  & =   2.237 \sigma \left(2.894 \Gamma_t + \sigma - \tau -\frac{2 \beta \left( 3 \Gamma_t +  \sigma  + \tau  - \beta \right) }{\tau + 2 \beta - 0.003275} \right) +0.45 \tau - 2 \beta + 0.026
    \end{aligned}
\end{equation*}
respectively, where $\tau = T-t$, as before.

\section{Conclusion}\label{sec:concl}

The paper demonstrates the merits of a GP approach to solving optimal hedging problems under transaction costs. This heuristic method delivers approximations that are analytic functions of various parameters describing the option contract, the dynamic of the underlying, and transaction costs. These explicit approximations to optimal hedging strategies can easily be tested and integrated into an automated trading system.

Other estimation methods, such as Zakamouline's (2006), derive closed-form approximations using ad hoc specifications of the functional form and an exact (numerical) solution for their calibration. In contrast to such approaches, ours is both simpler and more general, as it requires neither explicit solutions nor assumptions about functional forms. Preliminary results indicate that our GP method also works well for exotic options, other nonlinear transaction costs and stochastic volatility models but that is beyond the scope of this short note.

\section*{References}

\begin{list}{}%
{\leftmargin=2em \itemindent=-2em}

\item \c{C}etin, U. and Rogers, L.C.G., Modeling liquidity effects in discrete time, Mathematical Finance, 2007, \textbf{17}, 15--29.

\item Davis, M.H.A., Panas, V.G. and Zariphopoulou, T., European option pricing with transaction costs, SIAM Journal on Control and Optimization, 1993, \textbf{31}, 470--493.

\item Hodges, S.D. and Neuberger, A., Optimal replication of contingent claims under transaction costs, Review of Futures Markets, 1989, \textbf{8}, 222--239.

\item Koza, J.R., Genetic Programming: On the Programming of Computers by Means of Natural Selection, MIT Press, 1992.

\item Malo, P. and Pennanen, T., Reduced form modeling of limit order markets, Quantitative Finance, 2012, \textbf{12}, 1025--1036.

\item Monoyios, M., Option pricing with transaction costs using a Markov chain approximation, Journal of Economic Dynamics and Control, 2004, \textbf{28}, 889--913.

\item Nordin, P., Evolutionary Program Induction of Binary Machine Code and its Applications, Krehl Verlag, 1997.

\item Whalley, A.E. and Wilmott, P., An asymptotic analysis of an optimal hedging model for option pricing with transaction costs, Mathematical Finance, 1997, \textbf{7}, 307--324.

\item Zakamouline, V.I., Efficient analytic approximation of the optimal hedging strategy for a European call option with transaction costs, Quantitative Finance, 2006, \textbf{6}, 435--445.

\end{list}

\end{document}